\documentclass{elsart}
\usepackage{hhline}
\usepackage{graphicx}
\usepackage{amssymb}

\begin{document}
\begin{frontmatter}

\title{Trap Activation Energy and Transport Parameters of HgI$_2$ Crystals for Bubble-Plasma Diagnostics.}

\author[IPTP]{M.B. Miller}\footnote{mmiller@jinr.ru},
\author[FLNR JINR]{V.F. Kushniruk},
\author[IPTP]{A.V. Sermyagin},
\address[IPTP]{Institute of Physical and Technology Problems,
141980, Dubna, Russia}
\address[FLNR JINR]{Flerov Laboratory of Nuclear Reactions,
JINR, 141980, Dubna, Russia}
\maketitle
\begin{abstract}
In recent data on neutron induced acoustic cavitation in
deuterium--containing liquids obtained by neutron measurements it
was shown that very high temperatures could arise in some special
cases. To study temperature of so--called bubble plasma it is
desirable to have micro--detectors of X--rays, which can be
prepared on the basis of room--temperature semiconductor
detectors, in particular on mercuric iodide ($\alpha$--HgI$_2$)
crystals. Having in view this aim, the properties of gel--grown
($\alpha$--HgI$_2$) crystals was studied by means of isothermal
currents, and trap parameters was estimated. Results are promising
for special aim of preparing X--ray detectors with moderate energy
resolution needed in bubble--plasma diagnostic, though improving
of crystal growing technology is necessary\footnote{Two authors
(M.M.$\&$A.S.) are grateful to Russian Foundation for Basic
Research for financial support of this work,
Pr.\,\#02--02--16397--a.}.


\end{abstract}
\end{frontmatter}
\parindent=1cm
Recent indication on high temperature plasma arising in collapse
of cavitation micro--bubbles is of high importance~\cite{1}. The
data was obtained by neutron measurements in deuterium-containing
liquid, and they have only quality character in what regards
plasma temperature. For obtaining the value of temperature it is
desirable to measure the $X$--rays from the bubbles, for which aim
it is necessary to have micro--detectors. Such type of instruments
can be prepared on the basis of room--temperature semiconductor
detectors, in particular on mercuric iodide ($\alpha$--HgI$_2$)
crystals.

Mercuric iodide ($\alpha$--HgI$_2$) is a wide bandgap
($E_g=2.14$~MeV) high atomic number ($\langle Z\rangle=62$)
semiconductor, which has attracted considerable interest as a low
noise, room temperature $X$--ray and gamma ray detector. The
detection properties are strongly influenced by energy levels in
the forbidden gap arising from chemical impurities and/or
structural defects. In the present investigation we carried out
measuring isothermal currents on gel--grown HgI$_2$ platelets, in
order to estimate trap parameters. The deep energy level $\Delta E_t$ can be
determined using the following equation~\cite{2,3}:
\begin{equation}\label{1}
\Delta E_t=kT\ln(N_c\sigma_t\nu_{th}\tau_D),
\end{equation}
where\\$k$ -- Boltzman's constant,\\ $T$ -- absolute
temperature,\\ $N_c$ -- effective density of states in the
conduction band,\\ $\sigma_t$ -- trapping cross section,\\
$\nu_{th}$ --  thermal velocity of electrons,\\ $\tau_D$ --
detrapping time constant.\\

Detrapping time and corresponding activation energy of traps are
summarized in the Table~\ref{T}. Moreover, the estimate of
mobility--lifetime product of the electrons $(\mu\tau)_e$  was
performed by measuring the pulse height from alpha--particles
$^{239}$Pu ($E_\alpha \sim 5.15$\,MeV). We estimated
$(\mu\tau)_e\cong(1\div5)\cdot 10^{-6}$\,cm$^2$/V. This value is
about one order of magnitude lower than in the best detector grade
materials. In future, it would be of interest to study the nature
of defects observed and to improve the technology of growing
crystals. Taking into consideration not extremely high energy
resolution necessary for the aims of diagnostic micro--bubble
plasma the above results can be considered as encouraging. {
\begin{center}
\begin{table}[t]
\caption{}\label{T}
\begin{tabular}{||c|c|c||}\hline
\multicolumn{3}{||c||}{\begin{tabular}{c}Detrapping times and trap
activation energies\end{tabular}}\\\hline
\begin{tabular}{c}Crystal\\number\end{tabular}&\begin{tabular}{c}$\tau_D$,\\ $[$s$]$ \end{tabular}&\begin{tabular}{c}$E_t\pm 0.01${,}\\ $[$eV$]$ \end{tabular}\\
\hline
\begin{tabular}{c}1\\\\\\\end{tabular}& \begin{tabular}{c}185\\1110\\1860\\\end{tabular} & \begin{tabular}{c}0.82\\0.86\\0.90 \\\end{tabular} \\\hline
\begin{tabular}{c}2\\\\\\\\\end{tabular}& \begin{tabular}{c}50\\210\\1560\\5940 \\\end{tabular} & \begin{tabular}{c}0.78\\0.82\\0.87\\0.91 \\\end{tabular} \\\hline
\begin{tabular}{c}3\\\\\\\\\end{tabular}& \begin{tabular}{c}43\\225\\1470\\5100 \\\end{tabular} & \begin{tabular}{c}0.78\\0.82\\0.87\\0.90\\\end{tabular} \\\hline
\begin{tabular}{c}4\\\\\\\\\end{tabular}& \begin{tabular}{c}180\\1680\\2940\\7800\\\end{tabular} & \begin{tabular}{c}0.82\\0.87\\0.89\\0.91\\\end{tabular} \\\hline
\end{tabular}
\end{table}
\end{center}}

\end{document}